\documentclass[proceedings,conferenceproceedings,submit,oneauthor,pdftex,10pt,a4paper]{mdpi} 

\usepackage{graphicx}
\usepackage{siunitx}
\usepackage{inputenc}

\newcommand{\jetRAA}{$R^\mathrm{jet}_\mathrm{AA}$}
\newcommand{\jetRsA}{$R^\mathrm{jet}_\mathrm{sA}$}

\newcommand{\Ncoll}{$N_\mathrm{coll}$}
\newcommand{\pT}{$p_\mathrm{T}$}
\newcommand{\ET}{$E_\mathrm{T}^\mathrm{trig}$}

\newcommand{\EA}{$\mathrm{EA}_\mathrm{est}$}

\firstpage{1} 
\pubvolume{xx}
\issuenum{1}
\articlenumber{5}
\pubyear{2018}
\copyrightyear{2018}
\history{Received: date; Accepted: date; Published: date}

\Title{Event activity measurements and mid-rapidity correlations in 200 GeV p+Au collisions at STAR$^{\dagger}$}

\Author{David Stewart $^{1}$\orcidA{} for the STAR Collaboration}

\AuthorNames{David Stewart}

\address{%
$^{1}$ \quad Yale University; david.j.stewart@yale.edu
}

\firstnote{Presented at Hot Quarks 2018 - Workshop for young scientists on the physics of ultrarelativistic nucleus-nucleus collisions, Texel, The Netherlands, September 7-14 2018}

\abstract{
    These proceedings report preliminary measurements of correlations between
    mid-rapidity charged tracks and high-rapidity event activity (EA) at
    STAR for $\sqrt{s_\mathrm{NN}}=\SI{200}{GeV}$ p+Au collisions taken in
    2015. These correlations are intriguing because they inform the current debate
    over use of the Glauber model in ``small'' systems (here meaning p+A or d+A
    and denoted as ``s+A'') and have implications for calculating nuclear
    modification and quenching observables in these systems. 
    The results support concerns about centrality binning in p+Au collisions,
    and as such motivate using ratios of semi-inclusive, as opposed to fully
    inclusive, jet spectra to look for jet enhancement or suppression.
}


\begin{document}

\section{Introduction}

The statistical distributions of binary nucleon-nucleon collisions in A+A
collisions (\Ncoll{}), calculated by the Glauber Model \cite{David:2017vvn}, as
a function of impact parameter $b$, play an important role in probing nuclear
modification in A+A collisions. The \Ncoll{} distributions are binned into
centrality classes. In each distribution, \Ncoll{} is maximum for the most central
bin, i.e. with $b\rightarrow0$, and decreases monotonically through the most
peripheral bin.
  The results of hard scatterings are compared in ratio to
those from p+p collisions scaled by \Ncoll{}. For jets, the deviation of this
ratio (\jetRAA{}) from unity is an indication of nuclear modification.  In A+A
events, the strong suppression of \jetRAA{}
is an indicator of quark gluon plasma (QGP) formation.  The traditional
assumption was that small systems would not form a QGP.  Therefore, \jetRsA{}
was measured in order to study cold nuclear matter effects.

The first inclusive measurements of \jetRsA{} were reported for \SI{5.02}{TeV}
p+Pb collisions at the LHC by ALICE \cite{Adam:2016jfp}, CMS
\cite{Khachatryan:2016xdg}, and ATLAS \cite{ATLAS:2014cpa}, and for
\SI{200}{GeV} d+Au collisions at PHENIX at RHIC \cite{Adare:2015gla}.  As
expected, the values of \jetRsA{}, when not binned into centrality classes,
were consistent with unity.  However, centrality binned \jetRsA{} showed
significant suppression/enhancement for central/peripheral collisions at both
ATLAS and PHENIX, a similar result to that interpreted as a QGP signal in A+A
collisions.  Tantalizingly, this coincided with larger community interest in
small systems as a variety of particle collectivity signals were observed in
s+A collisions.

\section{Event activity estimation and correlations to \jetRsA{}}

Calculating \jetRAA{} assumes that the probability of a hard scattering scales
linearly with \Ncoll{}, which is applied to collisions by assuming that it
scales monotonically with a measured event activity estimation (\EA{}).  In the
above measurements, \EA{} values were determined by detectors at $\eta$ values
outside of the region where the jets were reconstructed in order to avoid
auto-correlations between \EA{} and \jetRsA{}.

The observed suppression/enhancement of \jetRsA{} may artificially result from
difficulties applying \Ncoll{} which are unique to small systems.  First,
compared to A+A collisions, small systems have large fluctuations in \EA{}
coupled with a relatively limited range of \Ncoll{}.  This can result in a
dynamical bias when calculating \jetRsA{} \cite{Adam:2014qja}. More intriguing
is the possibility that individual nucleon-nucleon collisions within a single
s+A collision that share a common nucleon are not independent. The effects of
such a correlation would be strongly evident in a p+A collision relative to an
A+A collision. In the former, every nucleon-nucleon collision shares the same
proton; in the latter many independent sets of such collisions would be
superimposed, thereby masking the effects of the correlation.

A study of p+Pb collisions at LHC energies concluded that a 20\% suppression of
soft particles correlated with the presence of a hard scattering would
reproduce the enhancement/suppression observed in \jetRsA{}
\cite{Bzdak:2014rca}. Energy conservation of a proton (or deuteron) common to a
set of nucleon-nucleon collisions may provide the physics mechanism for this
correlation.  Jet production would require a hard scattering in one
nucleon-nucleon collision, and therefore a reduction in energy available for
the production of soft particles in the remaining collisions.  Two such theory
calculations found this to be a sufficient explanation for the observed
suppression of central \jetRsA{}, one of which also found it sufficient for the
peripheral enhancement \cite{Armesto:2015kwa}\cite{Kordell:2016njg}.

\section{Correlations of \EA{} to mid-rapidity charged tracks at STAR for p+Au collisions}

\begin{figure}[h]
    \centering
    \includegraphics[width=34pc]{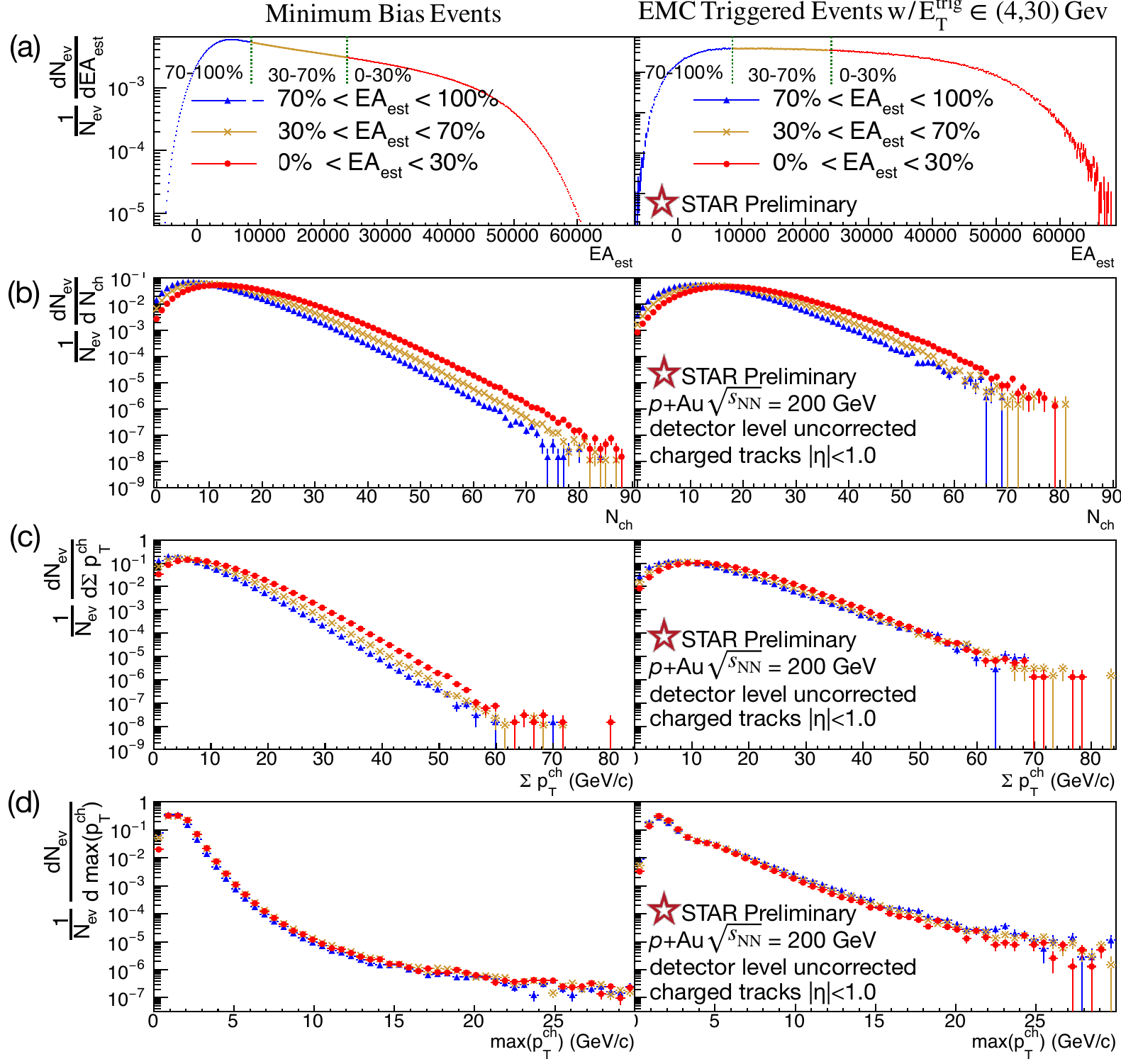}\hspace{2pc}%
    \caption{\label{STARdist} 
    Correlations between charged tracks at STAR ($|\eta|<1$) and \EA{}, the
    Au-going BBC inner ring signal ($|\eta|\in(3.3,5.8)$).  (a) left: \EA{}
    distribution of MB events, binned for the maximum and minimum 30\% as well
    as the middle 40\% of events.  (a) right: distribution of \EA{} in for
    EMC triggered events; the binning boundaries and labels are
    determined by the MB distribution. Columns (b), (c), and (d) give the
    distributions for the low, mid, and high \EA{} bins for charged track
    multiplicity, summed \pT{}, and maximum \pT{} respectively.}
\end{figure}

\begin{figure}[h]
    \centering
    \includegraphics[width=34pc]{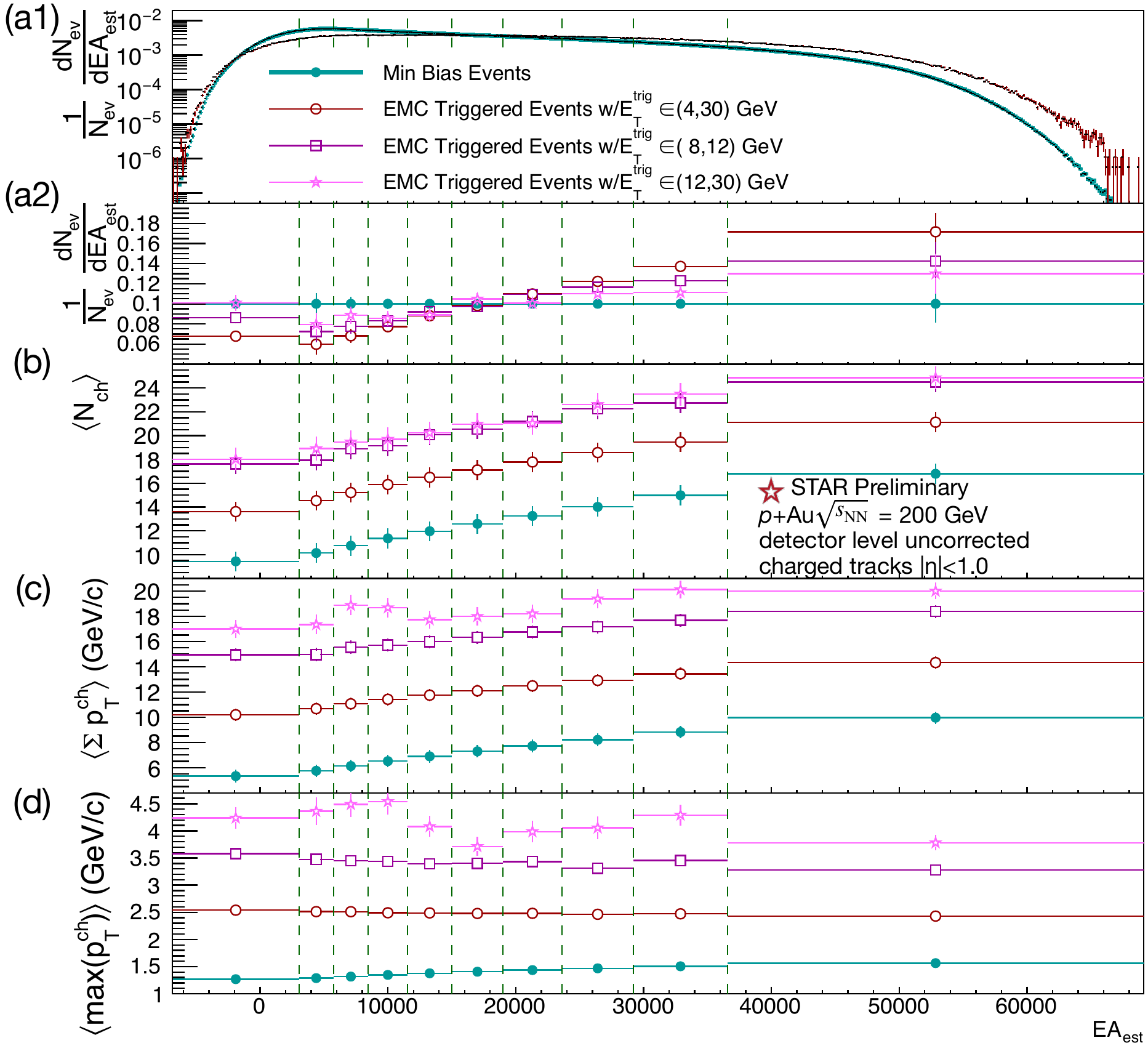}\hspace{2pc}%
    \caption{\label{STARresults} 
    Correlations between charged tracks at STAR ($|\eta|<1$) and \EA{}, the
    Au-going BBC inner ring signal ($|\eta|\in(3.3,5.8)$).  Each panel is
    plotted with data from MB events as well EMC \ET{} triggered events.  (a) The
    number of events of each BBC bin; each bin's boundaries are selected to
    contain 10\% of the MB data. (b) Average number of charged tracks per
    event. (c) Average sum of \pT{} per event. (d) Average maximum track \pT{}
    per event.} 
\end{figure}

STAR has a large set of $\sqrt{s_\mathrm{NN}}=\SI{200}{GeV}$ p+Au collisions
recorded in 2015 which will help address questions raised by the jet
measurements already released by PHENIX and the LHC experiments.  Measurements
of mid-rapidity charged track correlations to high-$\eta$ \EA{}, presented in
Figure~\ref{STARdist} and Figure~\ref{STARresults}, clearly indicate that
further study is required prior to calculating the \jetRsA{}.  

The figures present data from events with two separate triggers.  First:
minimum bias (MB) events.  Second: events triggered by the electromagnetic
calorimeter (EMC) selected by the hit with the maximum transverse energy
(\ET{}). 
The correlations reported are for charged tracks measured in the time
projection chamber (TPC) which has good track resolution from 0.2 to
\SI{30}{GeV/c} and has, as does the EMC, full azimuthal coverage. \EA{} is
measured as the sum of the signal in the inner ring of the beam beam counter
(BBC) in the Au-going direction. The BBC consists of sets of plastic
scintillators arrayed around the beam pipe; the inner ring of which covers
rapidity range 3.3 to 5.8.  

These preliminary results are detector level and uncorrected for detector acceptance 
or inefficiency effects. However, the
conclusions presented depend on the data's monotonicity and relative
distributions and are consequently not sensative to detector tracking efficiencies
and pileup. Statistical uncertainties are plotted for all data. Additionally,
a small relative trigger bias is added in quadrature with the statistical 
uncertainty in Figure~\ref{STARresults} (a2), (b), and (c). This bias
is quantified by the difference in results
from the MB data when cut for EMC hits so as to mimic the EMC trigger
and the results from the actual EMC triggered data.


Figure~\ref{STARdist} (a) shows the \EA{} distributions for MB events (on left)
and EMC triggered events (on right). As in the case of centrality in Glauber
calculations, \EA{} percentiles are defined by the MB distribution with 100\%
being the lowest and 0\% the highest value. Here the EA distribution is divided
into low (100-70\%), medium (30-70\%), and high (0-30\%) activity bins.  For
each \EA{} bin, the distribution of multiplicity, summed \pT{}, and maximum
single track \pT{}, are given in (b), (c), and (d).  Compared to A+A
collisions, the distributions in p+Au collisions heavily overalp among \EA{}
bins.  As expected, the mean values of multiplicity and summed track \pT{} are
higher for the high \EA{} bins. The normalization of the curves in (d) is
dominated by the first few, low-\pT{}, bins. In these bins the spectra from
high, medium, and low \EA{} events are fairly comparable. After these low \pT{}
bins, the spectrum of each \EA{} bin continues to be roughly equivalent for
harder scatterings in the MB data;  however, for the EMC triggered data, the high
\EA{} bin's spectrum is somewhat suppressed relative to the low \EA{} bin's
spectrum.

Figure~\ref{STARresults} (a1) gives the same \EA{} distributions (MB and EMC
triggered) shown in Figure~\ref{STARdist} (a). The \EA{} binning, consistent
between all plots, is selected for uniform numbers of MB events.
%
%
The remaining panels plot average (a2) (normalized) number of events, (b)
multiplicity, (c) summed track \pT{}, and (d) maximum single track \pT{}. The
results of two subsets of the EMC data, with higher trigger thresholds, are
also plotted.  The most striking results are deviations from what would result
if \EA{} and \Ncoll{} scaled positively, monotonically, together. If true, that
scaling would result in correlations in (a2)-(d) that are: (1) positive,
except, of course, for MB events in (a2), and, (2) smallest for MB events and
successively larger for EMC triggered events with successively higher trigger
thresholds.  Instead, while each correlation is positive in (a2), the correlations are smaller
for each successively higher \ET{} threshold; the data is too limited for
harder triggers to see if there would be an actual turnover at a sufficiently high threshold.
Correlations in (c) and (d) are about as naively expected, although it
is curious that mean multiplicity and summed \pT{} appear relatively saturated
by the time there is an \SI{8}{GeV} \ET{} such that they increase only slightly
with the higher \SI{12}{GeV} threshold.  Most notably, each
EMC triggered distribution (d) is anti-correlated; this directly
contradicts the assumption that hard scatterings (and therefore naively higher
\ET{} values) scale linearly with \EA{}.

\section{Summary}

Correlations of mid rapidity charged track observables to high $\eta$ \EA{}
have been presented. The results imply challenges to the use of \EA{} to
determine associated \Ncoll{} values. These difficulties motivate the use of
semi-inclusive jet spectra, in which \Ncoll{} cancels in the
ratio, instead of fully inclusive spectra, to probe for
jet suppression/enhancement in p+Au collisions \cite{Acharya:2017okq}\cite{Adamczyk:2017yhe}.

\vspace{6pt} 

\funding{This research was funded by U.S. Department of Energy under grant number  DE-SC004168.}

\reftitle{References}

\externalbibliography{yes}
\bibliography{inspire}

\end{document}